# SYSTEMATIC REVIEW ON PROJECT ACTUALITY

Suzana Cândido de Barros Sampaio [1,2], Marcelo Marinho[1,2], Hermano Moura[1]

[1]Informatics Center (CIn), Federal University of Pernambuco (UFPE)
[2]Department of Statistics and Informatics, Federal Rural University of Pernambuco (UFRPE), Brazil
Recife, PE, Brazil

*ABSTRACT*

*Nowadays much is written about how to manage projects, but too little on what really happens in project actuality. Project Actuality came out in the Rethinking Project Management (RPM) agenda in 2006 and it aims at understanding what really happens at project context. To be able to understand project actuality phenomenon, we first need to get a better comprehension on its definition and discover how to observe it and analyse it. This paper presents the results of the systematic review conducted to collect evidence on Project Actuality. The research focused on four search engines, in publications from 1994 to 2013. Among others, the study concludes that project actuality has been analysed by several methods and techniques, mostly on large organization and public sectors, in Northern Europe. The most common definitions, techniques, and tips were identified as well as the intent of transforming the results in knowledge.*

## 1. INTRODUCTION

This study aimed on identifying the theory associated with project actuality phenomenon as well as to analyse how academics and organizations are analysing it. It was also our concern which methods and techniques are being used and carried on in which size of organizations and through which kind of studies. To attend our goal, a systematic review was conducted by two researchers and PhD candidates from a post-graduate Program at Informatics Center (CIn) at Federal University of Pernambuco (UFPE), in Brazil. A third researcher was planned, but ended up helping only as triggered punctually just in case of disagreements, as a moderator.

A systematic review is a defined way of identifying, assessing, and analysing published primary studies in order to investigate a specific research question [1], [2]. It provides the means to identify, evaluate and interpret all available research relevant to a particular research question, topic area, and phenomenon of interest [3], [4]. It is a planned and ideally repeatable way of synthesizing results from the existing body of scientific literature. A systematic review follows a formal protocol to conduct research on a particular topic, with a well-defined sequence of methodological steps [5].

Actuality research argues that while a great deal is written about traditional project management, we know very little about the reality of project based on working and management itself [6]. Project actuality was one of the themes discussed in a research network – Rethinking Project Management – were the main argument was to enrich and to extend the subject of project management beyond its current conceptual foundations [7].

In order to investigate and understand the "actuality" of project, it is necessary to analyse how researchers are observing and analysing the actuality of software project and its management. This study is part of a PhD research on designing an approach to analyse and understand project actuality. Therefore, it is necessary to identify the methods used on analysing project actuality, and better understanding of this phenomenon. The systematic review method would be useable for this purpose and its results are presented here.

We briefly present below the results of the identification, analysis, and synthesis of project actuality definitions, along with what and how academics and organizations are analysing it.

## 2. RESEARCH METHOD

This section describes the course of each step in the methodology used to carry out this systematic review. The application of a systematic review requires a well-defined and sequential set of steps according to an appropriately designed study protocol. We followed Kitchenham's methodological guideline [1] for systematic reviews. The following subsections describes details on the course of these steps.

### 2.1 The environment

Before conducting the searches, a directory in the cloud was created with access to all researchers. A free web store service was used by all researchers to store all articles identified, as well as datasheets, partial reports and others. This enabled a full control of artefacts by the researchers, enabling each researcher to access the artefacts as if in a local environment, even thought they were remote. We also developed some datasheets to be used in all phases. The datasheets facilitated the organization and the standardization of data in many aspects, such as publications searched, study, size, filters to extract objective information, and more. As we had only two active researchers, both had to work as readers, analysers and synthesizers. We used Cin-UFPE labs to conduct all searches.

### 2.2 Search strategy

The strategy for a systematic review is a search plan to identify and regain the smallest publication's set that meet the systematic review criteria. The criteria are conditions to define if primary studies addresses the research questions defined for the systematic review [1]. The search strategy results in a protocol, which defines the research goal and procedures that includes the list of database engines and its research strings, research questions, selection criteria and steps to be conducted along the review, extraction procedures among others.

The research objective were to analyse project actuality phenomenon, with the purpose of consolidating the concept of actuality and identify the methods current used to observe and analyse project actuality, and analyse how to improve the management of projects based on these findings.

Search data engines were used to ensure unique results through the search for the same set of keywords. Our selection criteria were papers that contained an explicit statement about project actuality. For this research we selected only publications written in English. This choice was to its adoption by most international conferences and journals related to project management and for being the language used by most publishers related to the topic listed in the CAPES Journals Portal.

Before engaging in the real search, an initial study was conducted for all phases, denominated "pilot study". The pilots were performed to align phase to phase the understanding among researchers and to test all search engines. The study only proceeded after all researchers are in agreement with its results.

This review covered the period from 1994 to 2013 in four research database engines: ACM Digital Library, IEEEXplore, Science Direct and Springer Link. The engines were divided among the researchers.

## 2.3 The Search

According to Kitchenham [1] the aim of a systematic review is to find as many primary studies relating to the research question as possible using an unbiased search strategy. The search terms used this study were developed using the following steps [1], [3]:

1. Derive major search terms from the research questions by identifying Population, Intervention, Outcome, and Context.
2. Identify keywords and terms in the relevant papers.
3. Use Boolean OR to construct search strings from the search terms with similar meanings.
4. Use Boolean AND to concatenate the search terms and restrict the research. Should still be considered the publications:

The resulting search string was 'project actuality' OR 'actuality of project' or (actuality and 'software project').

## 2.4 The Selection

The idealised selection process has three step: the automatic selection (the search); selection stage one that cover the analyses of the title and abstract; and selection stage two that covers the analyses of the introduction and conclusion.

### 2.4.1 Automated Selection

The automated selection consists in the use of the research string in the chosen engines. In this phase, each researcher was responsible to find results in some engine and catalogued it following a designed template for it, where 800 papers were retrieved.

### 2.4.2 Selection Stage 1 – Title and Abstract

This phase consists in the selection of publications from the automatic search result set that could plausibly satisfy the selection criteria, based on reading 800 publications's title and abstract. Each researcher read the title and abstract of all 800 papers independently, and selected or excluded the publication. Jointly, researchers discussed their results.

Most of the papers, from the 800, mentioned actuality as a "time period", or as an expression, that means "nowadays" or "these days". The expression was never as keywords, and most of the times that it occurred in the abstract, meant something else. In this phase, if any disagreement, or doubt, the paper was included. This selection stage resulted in 36 publications that attend the criteria.

### 2.4.3 Selection Stage 2 – Introduction and Conclusion

The final selection stage consists in, from the selected list of papers that satisfy the selection criteria during selection stage 1, reading the introduction and conclusion of each paper. Once again, both researchers did all the reading independently, to reduce potential bias. No disagreement was found in this phase. Then, 19 publications were selected to be read during the extraction data phase.

## 2.5 Study Quality Assessment

In the Data Extraction and Quality Assessment, both researchers read a full paper for the quotes extraction, at the same time they did the methodological quality assessment of each publication. Both researchers performed the quality assessment. For each paper, it was analysed if the publication mention its conclusion (2), if it mentions as in conclusion (1), or as an ongoing investigation (0,5). Two other factors were assessed as follows, and were each marked YES or NO.

- Does the publication mention the possibility of selection, publication, or experimenter bias?
- Does the publication mention possible threats to internal or external validity?

The quality assessment was based only on whether the publication explicitly mentioned these issues. We did not make judgements about whether the publication had a "good" treatment of these issues. The results did not exclude any publication.

None of the studies mentions its full conclusion, ten of them mentioned as in conclusion and three entitled themselves as an ongoing investigation. The others did not mention anything about the status of conclusion.

## 2.6 Data extraction

This phase consisted in reading the entire publication and extracting quotes that answers on of the research questions. Before the beginning of this phase, we performed a new pilot to calibrate this conception. The researchers selected one random paper among the 19 publications selected. The entire paper was read and the extraction was conducted in parallel and the information were structured according to the data model and datasheet model created in the strategy plan, as the researchers discussed the results. Adjustments were necessary in the datasheet, and our final data model is shown in Figure 1.

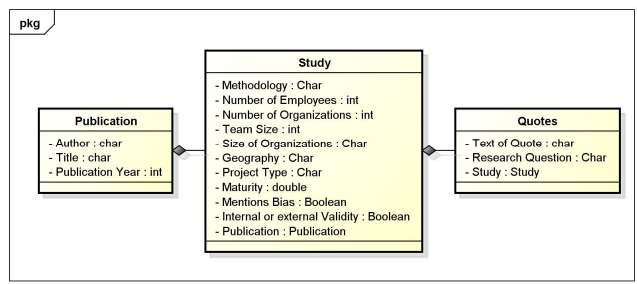

Fig 1. Research Data Model

A *Publication* is a technical report, conference article or journal article. From the 19 selected publications, the researchers identified six that did not showed relevant possible quotes to be extracted, resulting in 13 publications.

Besides all the presented criteria, the most cited publications in the references were considered relevant despite not having been identified by the automatic search. One new publication got into this selection and it was not indexed from the chosen engines, but it was cited by most of the publications selected. Which resulted in 14 final publications?

A *Study* in a publication describes an empirical investigation about organizations or researchers that have analysed project actuality or knowledge creation on software project actuality. Four publications did not mention the investigation method, but it was considered as a study for its relevance on developing a better understanding project actuality or ways to observe and analyse it. For each study, we extracted information about the attributes defined in datasheet.

Among the methodology on the 17 studies, the most common was Case Studies, 6 of them on a single organization and one with 3 organizations, followed by 4 Ethnographic Studies of one organization and others such as ANT (Actor–Network Theory), Grounded Theory, Pragmatic inquiry and Questionnaire.

From each study, we extracted a list of quotes. A *Quote* is a piece of text from the publication presenting the evidence that answer one of following research question:

- (RQ1) What is project actuality?
- (RQ2) What can be observed about project actuality?
- (RQ3) How can we observe and analyse project actuality?
- (RQ4) How current theories, concepts and methodologies underpinning project management research could be enriched by understanding project actuality?
- (RQ5) By understanding project actuality, can we enrich the knowledge created in the research process in project environments?

The main objective of this systematic review was to answer the first three questions. Anyways two extra questions were pointed out to maximize the analysis on project actuality phenomenon.
The two researcher worked separately in all the publications and by the end of the extraction phase, they agreed to an end set of 244 quote extracted from the 16 studies. Moreover, which question was answered by each quote. As mentioned before, six publications did not presented relevant quotes and they were excluded after this phase.

**2.7 Data Synthesis**

At the end of the Data Extraction phase, we had extracted 244 quotes, each containing answers for one of the five research questions. Once again, both researchers worked in the synthesis that consisted in three steps:

- Generate combinations of quotes and research questions;
- Independently reviewed each quote-question;
- Discussing the findings;
- In other to facilitate the analyses, the two researchers also agreed on grouping the quotes by similarities, such as techniques in RQ3.

There was a good level of agreement, differences in opinion were discussed in a meeting and they were easily resolved without the need of involving a third researcher arbitrating, as planned.
Some publications had quotes that answered more than one research question, as exhibit in Table 1. Research question 3 was the one with the most number of publications associated. It was also the question by far with most answers, with 154 quotes associated, as presented in Table 2. However, the first question, related to the phenomenon, the definition of project actuality as found in only fourteen quotes of six studies, as shown in Table 1 and 2.

International Journal of Computer Science & Information Technology (IJCSIT) Vol 6, No 5, October 2014

Table 1. Research Questions X Publications.

| Research Question | Number of Publications |
|---|---|
| RQ1 | 6 |
| RQ2 | 10 |
| RQ3 | 14 |
| RQ4 | 7 |
| RQ5 | 8 |

Table 2. Research Questions X Number of Quotes.

| Research Question | Number of Studies |
|---|---|
| RQ1 | 14 |
| RQ2 | 50 |
| RQ3 | 126 |
| RQ4 | 18 |
| RQ5 | 36 |

## 3. RESULTS

This section discusses our systematic review, presents an overview of the process and analyses how the results helps on answering ours research questions, as well as summarizes and discusses our findings.

### 3.1 Overview and process

In the automated selection phase, the searches were conducted in four sources. The Figure 2 shows the results obtained on each stage at systematic review process. At the synthesis phase, we have the number of quotes associated with each research question.

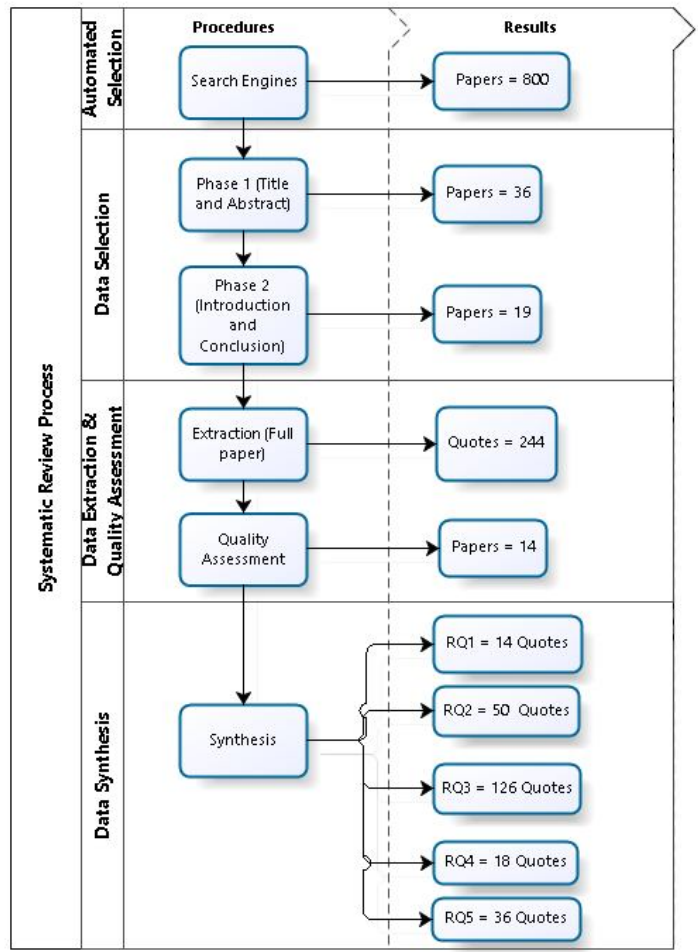

Figure 2 Results obtained on each phase of the systematic review process.

The Table 3 illustrates the amount of publications identified during each researchers phase, performed on each database engine. As it shows, at the end set of publications, no IEEEXplore was left to be extracted. Ten out of the thirteen papers selected were brought by only one search engine, the Science Direct.

Table 3.Number of publications gathered by engine in all phases.

| Phase- Activity | Number of Papers | Science Direct | Springer Link | IEEEXplore | ACM |
|---|---|---|---|---|---|
| Automated Selection | 800 | 42 | 47 | 23 | 688 |
| Selection Stage 1 | 36 | 12 | 11 | 6 | 7 |
| Selection Stage 2 | 19 | 11 | 2 | 5 | 1 |
| Extraction | 13 | 10 | 2 | 0 | 1 |

After performing the selection stage 1 and 2, that means after reading all introduction and conclusion off the selected paper, a total of 19 studies were selected. During the extraction phase, all 19 papers were read and studied. It lead to an exclusion of six publications that did not meet the inclusion criteria. At this phase one publication were included, for as a snowball analyses and

57

because it was cited by most of the publications. Table 4 gives the end list of selected studies, by year. Excluding only the last publication included.

Table 4. End list of publications per year.

| #ID | Authors | Publication Year |
|---|---|---|
| s3[16] | Judith Segal | 2005 |
| s6[6] | Svetlana Cicmil, Terry Williams, Janice Thomas , Damian Hodgson | 2006 |
| s10[7] | Mark Winter,Charles Smith, Peter Morris, Svetlana Cicmil | 2006 |
| s12[8] | Svetlana Cicmil | 2006 |
| s14[12] | Lynn Crawford , Peter Morris, Janice Thomas, Mark Winter | 2006 |
| s13[12] | Damian Hodgson and Svetlana Cicmil | 2007 |
| s1[13] | Gabriela Avram, Liam J. Bannon, Anne Sheehan, Anders Sigfridsson, Daniel K. Sullivan | 2008 |
| s7[10] | Chris Sauer, Blaize Horner Reich | 2008 |
| s8[17] | Eva Maaninen-Olsson, Tomas Mullern | 2009 |
| s4[15] | Daniel Sage, Andrew Dainty, Naomi Brookes | 2010 |
| s5[9] | Pierre-Luc Lalonde, Mario Bourgault , Alain Findeli | 2010 |
| s9[18] | Terry Williams ,Ole Jonny Klakegg , Ole Morten Magnussen , Helene Glasspool | 2010 |
| s2[19] | Steve Adolph & Wendy Hall & Philippe Kruchten | 2011 |
| s11[10] | J. Pollack , K. Costello, S. Sankaran | 2013 |

One interesting finding was that most of the studies on project actuality (13 studies) were from Northern Europe as shown in Figure 3. Region from where first presented the expression "project actuality", in the Rethinking Project Management agenda (RPM) [6].

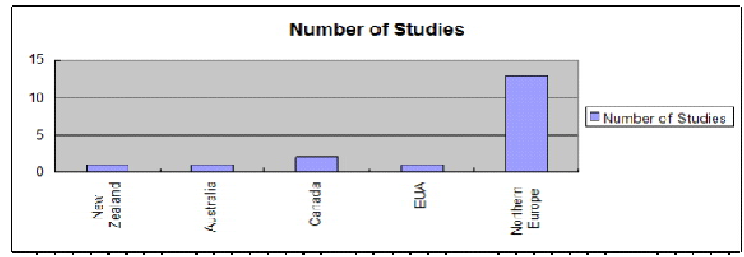

Figure 3.Number of studies per region-country.

Another finding was that most of the studies were in medium to large organizations. One study mentioned a 15 persons in a team (considered medium), another "medium to large", all the other did not mentioned or mentioned "large organization" (7 studies). Even the ones that did not mentioned it became easy to exclude micro or small organizations, by the cited characteristics. Besides one third of the studies were in public-sector.

### 3.2 Question 1 - What is project actuality?

Only six publications defined project actuality. Study 6 from Cicmil et al.[6] is the most cited publication and used definition. Most actors just adapted the definition presented by the authors or complemented it for a better understanding. For those who defined it, all of them have the same understanding actuality is a lived experience, focusing on social process and how practitioners think in action, which goes beyond the project itself, to encompass its relationships with other individuals, groups, organizations and concurrent projects.

Even though this question was more informative, it helped for a better understanding the phenomenon researched.

### 3.3 Question 2 - What can be observed about project actuality?

By understanding the first question, the second got easier to answer. We conclude that *what* is project context, relations and situations. Out of the ten publications that answered this question, Table 5 presents the most common aspect to be observed when dealing with project actuality.

Table 5. Most common "*what*".

| Technic/Strategy | Publications |
|---|---|
| Social process or social relations | s2, s5, s6, s7, s8, s12 |
| Context | s2, s5, s6, s8, s13 |
| Situation | s2, s5, s6, s12, s14 |
| Human action | s6, s7, s13, s14 |
| Relation | s5, s6, s8 |

The studies presented that to understand project actuality it is necessary to observe the social process, the context, actors' possibilities and options when confronted with a particular situation [9], "situating projects in existing and emerging social relations" [17][17], unusual things and activities rather than "universal" elements of perceived "good practice" [8].

Still on the *what* we have to observe: actors, meetings, central concern, critical movement, way to engage with projects, "what the actors do and how they do it", concepts of judgment, intention, motivation, purpose, value, reason for acting, problematic situations, speech, statement and related.

### 3.4 Question 3 - How can we observe and analyse project actuality?

For the third question, we grouped techniques to help answering the question. The quotes presented different tips, methods and techniques or strategies to analyse project actuality, such as observation, ethnographic methods, interview, building trust, documentation analysis, meetings follow up, to present the results to the interviewees, among others. Table 6 presents the most common techniques or strategies identified to observe and analyse project actuality.

Table 6. Most common techniques or strategies

| Technic/Strategy | Publications |
|---|---|
| Interview | s1, s2, s3, s5, s6, s7, s8, s9, s12 |
| Observation | s1, s2, s5, s8, s12, s13 |
| Documentation Analysis | s1, s2, s3, s5, s8, s9 |
| Meeting follow up | s1, s3, s5, s8 |

Besides the techniques, several tips and strategy were recidivists in the studies such as: observe meetings (4 quotes), attention on how to register and what to do with the data (9 quotes), recording Interviews (3 quotes), present the results to the interviewees (2 quotes), build trust (4 quotes), among others that just appear in one study.

### 3.5 Question 4 - How current theories, concepts and methodologies underpinning project management research could be enriched by understanding project actuality?

The studies present a flat understanding that by publishing the findings of project actuality we can generate a complementary set of knowledge that will help practitioners in project management practice. By understanding what really happens in project actuality, and better understanding of what literature to suggest or engage. Still, Cicmil et al. [6] has the publication with the most representative quote for the concept. Table 7 presents the most representative quotes.

Table 7. Most representative quotes associated with research question 4

| Publication | How to enrich project management theory? |
|---|---|
| s4 | The development of new forms of project management practice, training and education that are better able to apprehend social complexity, power relations and tacit knowledge and self-reflexive practice describe various acts that guide the inquiry process in project conduct and explore the bidirectional nature of the interactions[15]. |
| s6 | "Focused on serious consideration of 'knowledge in action', actuality research provides an insight into some shortcomings of the mainstream goal of disseminating 'best practice' in project management [6]. |
| s6 | Such a combination of theoretical and methodological approaches enabled researchers and participating practitioners to address together the important issues of project management praxis such as social responsibility, judgment, emotions, the operation of dominant discourses, power-knowledge relationship, and practical wisdom, which are rarely captured by conventional research methodologies in project management [6]. |

### 3.6 Question 5 - By understanding project actuality can we enrich the knowledge created in the research process in project environments?

Even though some may think this question is out of context, for those who study project actuality understand the relation. This is definitely a challenge, but doable. Table 8 presents the main idea to help the knowledge creation process, where reflectivity is crucial and it counts on a self-reflexive practitioner. One ready to engage in reflective activities. The quote from the study s14 represents it the most.

Table 8. Ways to enrich knowledge in project environment

| Publication | Way to enrich Knowledge in project environment |
|---|---|
| s4 | Turn to reflexivity as a synthesis or reconciliation of control/creativity[15]. |
| s5 | Describing various acts that guide the inquiry process in project conduct and explore the bidirectional nature of the interactions: descriptive practices support project-defining proposals, and design practices – expressed through discourse – convey the actors' intentions [9]. |
| s5 | We emphasize that this type of theorization can serve as a "tool of |
| | thought", a "cognitive tool" to nurture project management practice by systematically adopting a reflective and critical stance toward the inquiry trends [9]. |
| s6 | Actuality research provides an insight into some shortcomings of the mainstream goal of disseminating 'best practice'[6]. |
| s12 | By interpreting the empirical material gathered in the process of prolonged active interviewing and collaborative participative interpretation of accounts reflecting experiences of project practitioners, we can generate alternative understandings of what goes on in project practice [8]. |
| s14 | Less separation between learning and actual practice accompanied by less focus on knowledge acquisition and more emphasis on holistic capability development that extends beyond knowledge to encompass practical application and experience, attributes and behaviors [12][12]. |

## 4. LIMITATIONS AND VALIDITY THREATS

One of this study's highlights is that it is based on existing evidence in industry and science. Since these evidences were found in journals and conferences, it is expected that they have been evaluated against threats to validity. However, the research approach inherits the validity's threats from the studies listed in the publications found. In addition, despite the concern in using a rigorous methodological framework, this research has some limitations, such as:

- An analysis of the geographical distribution of studies and was performed, and most studies are concentrated in certain region, Western Europe. This is a threat to external validity, since it is not possible to know if the project actuality phenomenon are the same around the world.
- Due to constraints of time and budget, the survey did not consider some bases of data that are suggested by Kitchenham [3]: Wiley InterScience, InspecDirect, Scopus and Scirus. While this may represent a limitation and a threat to validity, the major conferences and journals were analysed during the manual search. The review protocol can be used to extend the results using these and other sources.
- Even though three researchers reviewed the search process, as suggested by the guides, most of the process was conducted just with two researchers. The third researcher was only a moderator or advisor, trigged as needed.

## 5. CONCLUSIONS

This paper presents the results of a systematic review on software project actuality. A total of 800 papers were returned, of which 19 were selected after the second stage of the selection. Later on 6 publications were discarded due to not answering any of the research questions. As a result, 13 studies were selected. During the extraction phase, 1 extra publication was added. These two came from an analysis of the 19 original selections citations. Data were extracted from all 14 studies and then synthesized against the defined research questions.

Our results showed that project actuality is not a solid phenomenon and not many studies were carried long after the RPM [6]. Cicmil is the most collaborative author in this subject. The concept is most known and analysed in Nortern Europe, by the public sector or medium to large organizations. For those who defined it all have the same understanding, that actuality is a lived experience, focusing on social process and how practitioners think in action, that goes beyond the project itself, to encompass its relationships with other individuals, groups, organizations and concurrent projects.

Besides, many methodologies were used to analyse it such as ANT (Actor–Network Theory), Case studies, Ethnography, Grounded Theory, Pragmatic inquiry, Questionnaire, observations and interviews. Independent of the method, interviews, observations, documentation analysis were the most flat technic used to observe and analyse project actuality.

The systematic review helped in a better understanding of project actuality and identified best ways to observe it and analyse it. Techniques and strategies found in this study were used and will be used to carry observations and analyses of software project in small organizations in Brazil.

## ACKNOWLEDGEMENTS

The authors would like to thank CAPES for supporting this research, as well as the Cin-UFPE for the labs availability to conduct all searches. We also would like to thank the reviewers for their contributions on this work.

**Authors**

Suzana Sampaio has 17 years of experience on the software development industry, working for the last 9 years as a consultant in process improvement, agile methodology and project management. She is currently a PhD candidate at Informatics' Center at Federal University of Pernambuco (CIn - UFPE). Her research focuses on Software Project Management where she investigates software project actuality. Also works as an assistant professor at Department of Informatics (DEINFO) at Federal Rural University of Pernambuco in Brazil (UFRPE). She holds a Masters in Computer Science (CAPES Concept 6) at University Federal of Pernambuco (UFPE), on Software Development Productivity (2010) and graduated in Computer Science (2000). She is an ISO 9001:2008, ISO 29110 Lead Auditor, MR-MPS-SW and CMMI implementer and appraiser. Besides, she has interest in the following lines of research: project management, agile project management, process improvement and quality assurance.

Marcelo Marinho is currently assistant professor in the Department of Informatics (DEINFO) of the Federal Rural University of Pernambuco (UFRPE). Conducts doctoral research at the Federal University of Pernambuco (UFPe) Center for Informatics (CIn) in the area of Project Management, where he investigates uncertainties in Software Projects. Holds a Masters in Computer Science (CAPES Concept 6) CIN by / UFPe. Graduated in Computer Science from the Catholic University of Pernambuco. He had been worked in the IT field for 10 years in which 6 in the area of project management software. Works in the area of Software Engineering and Information Systems, and developing research projects, mainly in the following lines: project management, project management software, software development process, IT management, software quality, software process and agile methodology.

Hermano Perrelli de Moura is an Associate Professor and Prorector for Planning and Finance, Federal University of Pernambuco (Recife, Brazil). PhD in Computing Science (1993), University of Glasgow, Scotland. MSc in Computing Science (1989), Federal University of Pernambuco, Brazil. Electronic Engineer degree (1982) from Federal University of Pernambuco, Brazil. PMP – Project Management Professional (2003) by Project Management Institute. Research areas of interest includes: management of software projects; software process improvement; strategic planning of information systems. Acted as a consultant on various projects involving definition and implementation of processes for project management and software development. Additional Information: Federal University of Pernambuco personal webpage: www.cin.ufpe.br/hermano. LinkedIn: http://www.linkedin.com/in/HermanoMoura .